\documentclass[final,1p,times]{elsarticle}
\usepackage{graphicx}
\usepackage{amssymb}
\usepackage{amsthm}
\usepackage[switch]{lineno}
\biboptions{numbers,sort&compress}

\newcommand{\pt}{$p_{T}$}

\journal{Nuclear Physics A}

\begin{document}

\begin{frontmatter}

\title{Beam energy dependence of strange hadron production from STAR at RHIC}

\author{Xiaoping Zhang (for the STAR\fnref{col1} Collaboration)}
\fntext[col1] {A list of members of the STAR Collaboration and
acknowledgements can be found at the end of this issue.}
\address{Department of Engineering Physics, Tsinghua University,
Beijing 100084, China}
\address{Space Science Institute, Macau University of Science and Technology, Macau, China}

\begin{abstract}
We present STAR measurements of $K^{0}_{S}$, $\phi$, $\Lambda$,
$\Xi$, and $\Omega$ at mid-rapidity from Au+Au collisions at
$\sqrt{s_{NN}}$ = 7.7, 11.5, 19.6, 27, and 39 GeV from the Beam
Energy Scan (BES) program at the BNL Relativistic Heavy Ion Collider
(RHIC). Nuclear modification factors and baryon-to-meson ratios are
measured to understand recombination and parton energy loss
mechanisms. Implications on partonic versus hadronic dynamics at low
beam energies are discussed.
\end{abstract}

\end{frontmatter} 


\section{Introduction}

The goals of the BES program at RHIC are to search for the Quantum
Chromodynamics (QCD) critical point as well as the phase boundary
between partonic and hadronic phases \cite{starbes,qcp}. Through
systematic study of Au+Au collisions at $\sqrt{s_{NN}} = 7.7 - 39$
GeV, one could access a wide region of temperature $T$ and baryon
chemical potential $\mu_{B}$ in the QCD phase diagram. Strange
hadron production is sensitive to parton dynamics in
nucleus-nucleus collisions. 
The enhanced production of strange hadrons in nucleus-nucleus
collisions with respect to proton-proton collisions at the same
energy has been suggested as a signature of Quark-Gluon Plasma (QGP)
formation in these collisions \cite{rafelski}. Up to now, strange
hadron yields have been extensively measured in many experiments at
different accelerator facilities \cite{e895, e896, e917, na49, na57,
starstrange, phenixla}. In these experiments, substantial
strangeness enhancement has indeed been observed, especially for
multi-strange hyperons.

Differential observable on strange hadron production at intermediate
to high transverse momentum ($p_{T}$) is also sensitive to the
degree of freedom of the system. At intermediate $p_{T}$ (2.5 $-$ 4
GeV/$c$), baryon-to-meson ratios in more central Au+Au collisions at
$\sqrt{s_{NN}}$ = 200 GeV are much higher compared to those in
elementary collisions \cite{strangercp, starpip, philong}. This is
consistent with the parton recombination picture for hadron
formation \cite{reconbination}. Such models require constituent
quarks to coalesce into hadrons, and hence the observed enhancement
of baryon-to-meson ratios at intermediate $p_{T}$ is an important
evidence for the formation of deconfined matter with partonic
degrees of freedom \cite{strangercp, starpip, philong}. At high
$p_{T}$ ($>4$ GeV/$c$), the suppression of hadron yields in central
Au+Au collisions at $\sqrt{s_{NN}} = 200$ GeV compared to peripheral
collisions is considered as a strong evidence of partonic energy
loss in the dense partonic medium \cite{strangercp}. Because of the
possible transition from partonic dominated phase to hadronic
dominated phase, it is expected that the established paradigm for
partonic degrees of freedom at top RHIC energy may break at a given
low collision energy. In order to locate the phase transition
boundary, it will be quite interesting to measure the energy
evolution of baryon-to-meson ratios and nuclear modification factors
($R_{\textrm{\tiny{CP}}}$, defined later).

With its large acceptance, STAR has collected high statistics Au+Au
data at $\sqrt{s_{NN}} = 7.7$, 11.5, and 39 GeV in 2010, and at
$\sqrt{s_{NN}} = 19.6$ and 27 GeV in 2011. This allows high
precision differential measurement of strange hadron production,
especially at intermediate to high $p_{T}$, at these energies.
Previous measurement on strange hadron production in $\sqrt{s_{NN}}
= 7.7$, 11.5 and 39 GeV has been reported in Refs. \cite{feng, zhu1,
zhu2, zhang1}. In this paper, we will present new experimental
results on strange hadron production in $\sqrt{s_{NN}} = 19.6$ and
27 GeV.

\section{Experimental data analysis}

The Au+Au collision events collected by the minimum bias trigger are
used in this analysis. STAR's Time Projection Chamber (TPC)
\cite{tpc} and Time-of-Flight (TOF) detector \cite{tof1} are used
for tracking and particle identification. The events are required to
have a primary $Z$ vertex (along beam direction) within $\pm70$ cm
from the center of the TPC for Au+Au collisions at
$\sqrt{s_{\tiny{\textrm{NN}}}} = 19.6$ and 27 GeV, to ensure nearly
uniform detector acceptance. After the event selection, we obtain
about 36 and 70 million Au+Au minimum bias triggered events at
$\sqrt{s_{\tiny{\textrm{NN}}}} = 19.6$ and 27 GeV, respectively. The
collision centrality is determined by the measured raw charged
hadron multiplicity from the TPC within a pseudorapidity window
$|\eta|<$ 0.5 \cite{glauber2}.

The TPC can identify charged particles for $p_{T} < 1$ GeV/$c$ by
their specific energy loss (d$E$/d$x$) while traversing the TPC gas.
With the fully installed TOF detector \cite{tof1} in the year 2010,
one can identify $\pi^{\pm}$, $K^{\pm}$, proton, and anti-proton for
\pt\ up to 3 GeV/$c$ statistically. The multi-strange hadron signals
and raw yields are obtained from the invariant mass distribution
reconstructed by their hadronic decay channels: $\phi\rightarrow
K^{+}+K^{-}$, $\Xi^{-}(\Xi^{+})\rightarrow \Lambda(\bar{\Lambda}) +
\pi^{-}(\pi^{+})$, and $\Omega^{-}(\Omega^{+})\rightarrow
\Lambda(\bar{\Lambda}) + K^{-}(K^{+})$. The decay daughters
$\Lambda(\bar{\Lambda})$ are reconstructed through
$\Lambda(\bar{\Lambda})\rightarrow p(\bar{p})+\pi^{-}(\pi^{+})$.
$K_{S}^{0}$ is reconstructed through its decay channel
$K_{S}^{0}\rightarrow \pi^{+}+\pi^{-}$. Detailed descriptions on
analysis cuts about decay topology, kinematics as well as particle
identification can be found in Refs. \cite{starstrange, philong,
feng, zhu1, zhu2, zhang1}.

After correcting the raw spectra for reconstruction efficiency and
geometrical acceptance, we obtain the corrected $p_{T}$ distribution
of $K^{0}_{S}$, $\phi$, $\Lambda(\bar{\Lambda})$,
$\Xi^{-}(\Xi^{+})$, and $\Omega^{-}(\Omega^{+})$ at mid-rapidity
($|y| < 0.5$) in different centralities for Au+Au collisions at
$\sqrt{s_{NN}} = 19.6$ and 27 GeV. The $\Lambda (\bar{\Lambda})$
spectra have been corrected for the feed-down contributions from
$\Xi^{-}(\Xi^{+})$ and $\Xi^{0}$($\bar{\Xi}^{0}$) weak decays
\cite{feng, zhu1, zhu2}.


\section{Results and discussions}
\begin{figure}[htbp]
\begin{center}
\includegraphics[width=1.0\textwidth]{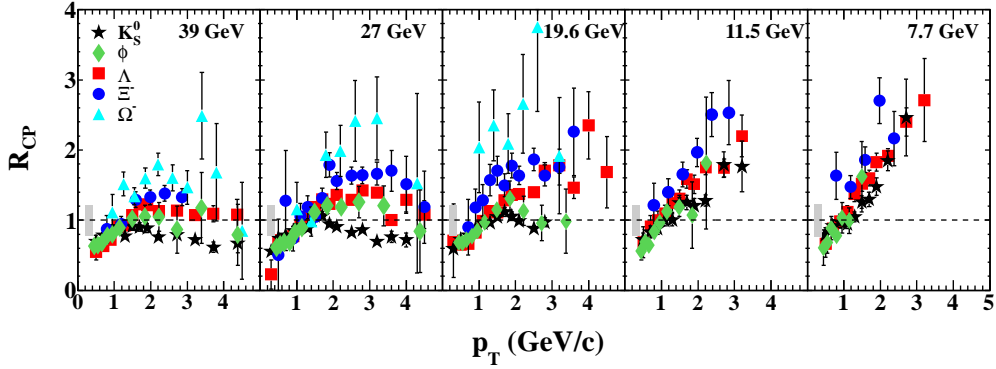}
\end{center}
\vspace{-4ex} \caption{(color online) $K^{0}_{S}$, $\Lambda$,
$\Xi^{-}$, and $\Omega^{-}$ $R_{\textrm{\tiny{CP
}}}$(0-5\%/40\%-60\%) at mid-rapidity ($|y|<0.5$) in Au+Au
$\sqrt{s_{NN}} = 7.7\ -\ $39 GeV collisions. $\phi$-meson
$R_{\textrm{\tiny{CP}}}$ is from 0-10\% and 40\%-60\% centralities.
Errors are statistical only. The left grey band is the normalization
error from $N_{\textrm{\scriptsize{bin}}}$. \label{figrcp}}
\end{figure}
In Fig. \ref{figrcp}, we present the $R_{\textrm{\tiny{CP}}}$ of
$K^{0}_{S}$, $\phi$, $\Lambda$, $\Xi^{-}$, and $\Omega^{-}$ in Au+Au
$\sqrt{s_{NN}} = 19.6$ and 27 GeV collisions. The
$R_{\textrm{\tiny{CP}}}$ data in 7.7, 11.5 and 39 GeV are from Refs.
\cite{zhu2}. The $R_{\textrm{\tiny{CP}}}$ is defined as the ratios
of particle yields in central collisions over those in peripheral
ones scaled by the number of inelastic binary collisions
$N_{\textrm{\scriptsize{bin}}}$.
Here, $N_{\textrm{\scriptsize{bin}}}$ is determined from Monte Carlo
Glauber model calculations \cite{glauber2}. The
$N_{\textrm{\scriptsize{bin}}}$ are $799.8\pm27.4$ ($841.5\pm28.4$),
$721.6\pm25.3$ ($761.1\pm26.7$), and $74.0\pm15.4$ ($82.0\pm18.4$)
for 0-5\%, 0-10\%, and 40\%-60\% Au+Au collisions at $\sqrt{s_{NN}}
= 19.6$ (27) GeV, respectively. The $R_{\textrm{\tiny{CP}}}$ will be
unity if nucleus-nucleus collisions are just simple superpositions
of nucleon-nucleon collisions. Deviation of these ratios from unity
would imply contributions from nuclear or medium effects. For \pt\
above 2 GeV/$c$, one can see from Fig. \ref{figrcp} that $K^{0}_{S}$
$R_{\textrm{\tiny{CP}}}$ increases with decreasing beam energies,
indicating that the partonic energy loss effect becomes less
important. While the cold nuclear matter effect (Cronin effect)
\cite{Cronin1} starts to take over at $\sqrt{s_{NN}} = 11.5$ and 7.7
GeV, and enhances the hadron yields at intermediate $p_{T}$. The
energy evolution of strange hadron $R_{\textrm{\scriptsize{CP}}}$
reflects the decreasing partonic effects with decreasing beam
energies. In addition, the particle $R_{\textrm{\tiny{CP}}}$
differences are apparent at $\sqrt{s_{NN}}\geq$ 19.6 GeV. However,
the differences becomes smaller at $\sqrt{s_{NN}} = 11.5$ and 7.7
GeV, which may suggest different properties of the system created in
$\sqrt{s_{NN}} = 11.5$ and 7.7 GeV Au+Au collisions compared to
those in $\sqrt{s_{NN}}\geq$ 19.6 GeV.

\begin{figure}[htbp]
\begin{center}
\includegraphics[width=0.55\textwidth]{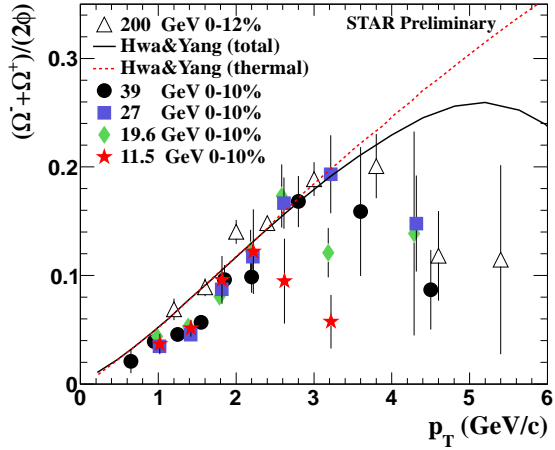}
\end{center}
\vspace{-3ex} \caption{(color online) The baryon-to-meson ratio,
($\Omega^{-}$+$\Omega^{+})/2\phi$, as a function of \pt\ in middle
rapidity ($|y| < 0.5$) from central Au+Au collisions at
$\sqrt{s_{NN}} = 11.5\ -$ 200 GeV. Errors are statistical only. The
solid and dashed lines represent recombination model calculations
for central collisions at $\sqrt{s_{NN}} = 200$ GeV
\cite{reconbination} with total and thermal strange quark
contributions, respectively. \label{figomphi}}
\end{figure}
Figure \ref{figomphi} shows the baryon-to-meson ratio,
($\Omega^{-}+\Omega^{+}$)/2$\phi$, as a function of \pt\ in Au+Au
central collisions at $\sqrt{s_{NN}} = 11.5$ $-$ 200 GeV. The 200
GeV data are from previous STAR measurements \cite{philong}. In
Au+Au central collisions at $\sqrt{s_{NN}} = 200$ GeV, the
intermediate-\pt\ (2.5 $-$ 4 GeV/$c$) $\Omega$ yield is explained by
mainly thermal strange quark recombination in the deconfined matter
\cite{reconbination}. The ($\Omega^{-}+\Omega^{+}$)/2$\phi$ ratios
in $\sqrt{s_{NN}} = 27$ and 39 GeV are close to that in 200 GeV.
However, the ($\Omega^{-}+\Omega^{+}$)/2$\phi$ ratios at
intermediate \pt\ at $\sqrt{s_{NN}}$ = 11.5 GeV are systematically
lower than those at $\sqrt{s_{NN}} \geq 27$ GeV. This may suggest a
change of strange quark dynamics between $\sqrt{s_{NN}} \geq 27$ GeV
and 11.5 GeV. Further study is needed to see whether this is caused
by the onset of deconfinement and/or phase space suppression of
multi-strange hadron production.

\section{Summary}

In summary, we have presented STAR measurements of $K^{0}_{S}$,
$\phi$, $\Lambda$, $\Xi$, and $\Omega$ at mid-rapidity from Au+Au
collisions at $\sqrt{s_{NN}} = 7.7$, 11.5, 19.6, 27, and 39 GeV from
the BES program at RHIC. Strange hadron $R_{\textrm{\tiny{CP}}}$
increases with decreasing beam energies, indicating that the
partonic energy loss effect becomes less important at low beam
energies. At intermediate \pt, particle $R_{\textrm{\tiny{CP}}}$
difference becomes smaller at $\sqrt{s_{NN}}$ = 7.7 and 11.5 GeV.
The intermediate-\pt\ ($\Omega^{-}+\Omega^{+}$)/2$\phi$ ratios at
$\sqrt{s_{NN}} = 11.5$ GeV are systematically lower than those at
$\sqrt{s_{NN}}\geq 27$ GeV, which may suggest a change of strange
quark dynamics between $\sqrt{s_{NN}} \geq 27$ GeV and 11.5 GeV.

\section*{Acknowledgments}
X. Zhang thanks the support by the National Natural Science
Foundation of China (Grant Nos. 10865004, 10905029, 11035009,
11065005, and 11105079), by the China Postdoctoral Science
Foundation (Grant No. 20100480017), and by the Foundation for the
Authors of National Excellent Doctoral Dissertation of P. R. China
(FANEDD) (No. 201021).


\end{document}